# Geometric transformation and three-dimensional hopping of Hopf solitons

Jung-Shen B. Tai[1], Jin-Sheng Wu[1] and Ivan I. Smalyukh[1,2,3*]

[1]Department of Physics and Chemical Physics Program, University of Colorado, Boulder, CO 80309, USA

[2]Department of Electrical, Computer, and Energy Engineering, Materials Science and Engineering Program and Soft Materials Research Center, University of Colorado, Boulder, CO 80309, USA

[3]Renewable and Sustainable Energy Institute, National Renewable Energy Laboratory and University of Colorado, Boulder, CO 80309, USA

* Correspondence to: ivan.smalyukh@colorado.edu

**Three-dimensional (3D) topological solitons are marvels of mathematical physics that arise in theoretical models in elementary particle and nuclear physics, condensed matter, and cosmology[1–4]. A particularly interesting type of them is described by the mathematical Hopf map from a hypersphere to an ordinary sphere[5], which in the physical 3D space exhibits inter-linked circle-like or knotted localized regions of constant order parameter values[2–4]. Despite their prevalence in models, such solitons remained experimentally elusive until recently, when hopfions were discovered in colloids[6] and chiral liquid crystals[7,8], whereas the so-called "heliknotons" were found both individually and within triclinic 3D lattices while smoothly embedded in a helical background of chiral liquid crystals[9]. Constrained by mathematical theorems[10,11], stability of these fragile 3D excitations is thought to rely on a delicate interplay of competing free energy contributions, requiring applied fields or confinement and impeding their practical utility[6,9]. Here we describe such 3D solitons in a material system where no applied fields or confinement are required for stability. Nevertheless, electric fields allow for inter-transforming heliknotons and hopfions to each**

**other, as well as for 3D hopping-like dynamics arising from nonreciprocal evolution[12] of the molecular alignment field in response to electric pulses. Stability of these solitons both in and out of equilibrium can be enhanced by tuning anisotropy of parameters that describe energetic costs of gradient components in the field, which is implemented through varying chemical composition of liquid crystal mixtures. Numerical modelling reproduces fine details of both the equilibrium structure and Hof-index-preserving out-of-equilibrium evolution of the molecular alignment field during switching and motions. Our findings may enable myriads of solitonic condensed matter phases[13] and active matter systems[14], as well as their technological applications.**

The discovery of Hopf fibration and topologically non-trivial hypersphere-to-sphere ($\mathbb{S}^3 \to \mathbb{S}^2$) mappings inspired the early proposals of continuous, spatially localized 3D topological excitations in diverse physical systems[2–5]. Among them, condensed matter systems with order parameters being unit-vector fields (order-parameter space $\mathbb{S}^2$), such as magnetization and electric polarization, or head-tail symmetric director fields (order-parameter space $\mathbb{S}^2/\mathbb{Z}_2$), such as molecular alignment in liquid crystals (LCs), are candidates for hosting Hopf solitons. The nontrivial nature of $\pi_3(\mathbb{S}^2)=\mathbb{Z}$ and $\pi_3(\mathbb{S}^2/\mathbb{Z}_2)=\mathbb{Z}$ groups informs about the existence of the corresponding topologically nontrivial constructs, but does not guarantee their stability. In fact, the Derrick-Hobart theorem predicts that such solitons cannot be stabilized within the simplest linear field theories[10,11]. By invoking chirality and a corresponding free energy term as the mechanism of overcoming the constraints of the Derrick-Hobart theorem in LCs and magnets[15], recent embodiments of stable Hopf solitons include hopfions hosted in a uniform background (constant order parameter far-field $\boldsymbol{n}(\boldsymbol{r}) \equiv \boldsymbol{n}_0$; Fig. 1a,e,h,k) and heliknotons in a helical background ($n(r)$) perpendicular to and twisting around the helical axis $\chi_0$; Fig. 1b,f,i,l)[6,7,16–20]. Hopf solitons were also modeled in conical backgrounds at externally applied fields, where $n(r)$ is at a cone angle $0° < \theta_c < 90°$ with respect to $\chi_0$[20] (Fig. 1c,g,j,m). In all cases, Hopf solitons classified by the third homotopy group exhibit interlinked regions of constant order parameter (preimages) with conserved linking number, identified with their integral topological charge – Hopf index $Q$ (Fig. 1e,f,g). Another notable feature of topology of Hopf solitons in elastically isotropic materials is that the streamlines of skyrmion number density $\Omega$ (or emergent magnetic field in magents[21,22]) also nontrivially link into Hopf fibrations[17,20], and the surfaces of constant skyrmion density form tori or handlebodies (Fig. 1h-j, Supplementary Video 1). This is of interest for novel 3D spintronic applications as the emergent fields describe the interaction between magnetic solitons and the spin currents[21,23]. Topological solitons attract fundamental as well as technological interest because of their topology-preserving multi-stability, field driven dynamics, and ability to act as individual particles or even form crystals[9,23–27]. Compared to their lower dimensional counterparts – 2D skyrmions, Hopf solitons have an advantage that they can be controlled in all three spatial directions.

Beyond chirality, the stability of Hopf solitons in experiments so far has also relied on geometrical confinement or externally applied fields. For example, hopfions in LCs, LC colloids, and magnets have been realized in thin film and nanodisk geometries where surface boundary conditions (BCs) were shown to be essential[6,7,16–19]. Heliknotons in LCs, on the other hand, can be stable in the bulk without confinement, but an external electric field was needed to sustain their stability[9]. While heliknotons in chiral magnets[20] and Hopf solitons as skyrmion knots in frustrated magnets[28] have been shown in numerical simulations to be stable *per se*, their experimental realizations are lacking. As a result, the 3D mobility and control of Hopf solitons is so far not fully utilized, precluding their technological applications. Moreover, though verified to be topologically identical, hopfions and heliknotons are distinct embodiments of Hopf solitons in their field configurations and are stable under different conditions. Insights into how their field geometries are related to each other are essential for understanding stability of Hopf solitons in and out of equilibrium. In conventional nematic LCs made of rod-like molecules, the twist deformation is favored over bend deformation[29], energetically hindering a smooth transition between the uniform state and the helical state through conical states, as well as the hopfion-heliknoton inter-transformation. However, recently, LCs made of bent-core molecules were found to have exceptionally low bend elastic constant within the nematic phase they form, and mixtures of rod-like and bent-core molecules exhibit tunable elastic anisotropy and are an ideal material system to explore stability of various Hopf soliton embodiments and their geometric inter-transformation[30,31].

In this work, we demonstrate structural stability of Hopf solitons under different elastic material constants, applied electric field $\bm{E}$, and confinement conditions in chiral LCs. Our numerical study reveals conditions when Hopf solitons are stabilized at no applied field or confinement in the helical background and when they are stabilized with the help of confinement or applied fields in the uniform or conical backgrounds. Furthermore, we identify a pathway for inter-transformation between hopfions and heliknotons, in confined LCs by switching $\bm{E}$, where $\bm{n}(\bm{r})$ transforms smoothly while preserving the soliton's topology. We experimentally demonstrate such facile inter-transformation. Our findings indicate that elastic constant anisotropy, tunable by adjusting the material composition, provides a new mechanism for enhancing stability of Hopf solitons in different backgrounds. Further, we have discovered a 3D hopping-like motion of Hopf solitons that arises from repeated inter-transformation between hopfions and heliknotons through periodic voltage switching.

We investigated the structural stability of elementary $Q = 1$ Hopf solitons in chiral LCs by minimizing the Frank-Oseen free energy density including the term describing dielectric effect of $\boldsymbol{E}$ [9,13,32]

$$f_{\text{CLC}} = f_{\text{elastic}} + f_{\text{electric}}$$

$$= \frac{K_{11}}{2}(\nabla \cdot \boldsymbol{n})^2 + \frac{K_{22}}{2}(\boldsymbol{n} \cdot \nabla \times \boldsymbol{n})^2 + \frac{K_{33}}{2}(\boldsymbol{n} \times \nabla \times \boldsymbol{n})^2 + \frac{2\pi K_{22}}{p_0}(\boldsymbol{n} \cdot \nabla \times \boldsymbol{n})$$

$$- \frac{\varepsilon_0 \varepsilon_a}{2}(\boldsymbol{E} \cdot \boldsymbol{n})^2 \qquad (1)$$

Here $\boldsymbol{n}(\boldsymbol{r})$ is the LC molecular alignment field; for continuous 3D solitonic excitations, it can be treated as a vector field for simplicity[6,7]. $K_{11}$, $K_{22}$, and $K_{33}$ are Frank elastic constants for bend, twist, and splay deformations, respectively, $p_0$ is the equilibrium pitch of the chiral LC at no field, $\varepsilon_0$ is the vacuum permittivity, $\varepsilon_a$ is the dielectric anisotropy of LC, and $\boldsymbol{E}$ is along $z$. Mimicking our experiments, we consider conditions where solitons can be stabilized in an unconfined bulk (Fig. 2a), for confinement with perpendicular BCs for $\boldsymbol{n}(\boldsymbol{r})$ along $z$ (Fig. 2b) and for unidirectional parallel BCs (Extended Data Fig. 3) by varying the elastic anisotropy associated with bend and twist deformation ($\sqrt{K_{33}/K_{22}}$) and the normalized electric field strength $\tilde{E}$ (Methods).

The translationally invariant backgrounds of $\boldsymbol{n}(\boldsymbol{r})$ for unconfined chiral LCs can be derived analytically by comparing their corresponding free energies (Methods), as shown with the help of $\theta_c$ contours in Fig. 2a. Hopf solitons can exist as stable or metastable particle-like, spatially-localized structures embedded in these different backgrounds (Fig. 2). Hopf soliton's $\boldsymbol{n}(\boldsymbol{r})$ configurations in the helical background (heliknotons) have energy density higher than the background when $\sqrt{K_{33}/K_{22}} \leq 1.2$ with $\tilde{E} = 0$ (Fig. 2a,e). At larger $\sqrt{K_{33}/K_{22}}$, heliknotons can remain metastable if $\tilde{E}$ is applied, consistent with previous findings that heliknotons required a stabilizing $\tilde{E}$ in chiral LCs with $\sqrt{K_{33}/K_{22}} > 1.5$ [9]. Hopf solitons can also be metastable in the bulk conical background for $30° \lesssim \theta_c \lesssim 40°$ at $\sqrt{K_{33}/K_{22}} \leq 0.4$ and $0.5 \leq \tilde{E} \leq 0.8$ (Fig. 2a,f). Excitingly, Hopf solitons are found to have $\boldsymbol{n}(\boldsymbol{r})$ with lower energy than their embedding backgrounds in parameter regions between those of metastable Hopf solitons and across the helical-uniform and helical-conical boundaries. In these parts of the diagram, the Hopf solitons fill the computational volume and form stable crystalline assemblies (Fig. 2a). Similar stable assemblies of Hopf solitons have been found in LCs and magnets, albeit here they can occur even without confinement or applied fields[9,20]. Importantly, regardless of the distinct embedding backgrounds with different $\theta_c$, the Hopf indices of the solitons remain unchanged, as evident from the linking number of preimages (Fig. 2d-f).

When confinement and perpendicular BC were applied to the chiral LC with a thickness $d = 3p_0$, the background $\theta_c$ contours shifted (Fig. 2b). The parameter regions of stability and metastability of Hopf solitons found in the bulk LC diagram are present also in the case of confinement. However, additional metastability regions of Hopf solitons emerge in the uniform background with $\theta_c = 90°$ (hopfions, Fig. 2d) under the confinement, differing from the case of no hopfions in the fully unwound LC bulk. This further demonstrates how confinement of LCs at surfaces with perpendicular BCs helps stabilize hopfions[6,7,17]. Remarkably, our structural stability diagram reveals that, with confinement and perpendicular BCs, hopfions and heliknotons can be metastable at different $\tilde{E}$ at $0.8 \lesssim \sqrt{K_{33}/K_{22}} \lesssim 1.3$, suggesting an *in situ* pathway for inter-transformation by varying the applied voltage. Beyond their structural stability region, Hopf solitons either collapse into the topologically trivial background through proliferation and annihilation of singular defects[16] or transform into point-defect-dressed solitonic structures such as torons[33] (Extended Data Figs. 1 and 2). Hopf solitons in confinement with parallel BCs show structural stability similar to that in bulk LCs (Extended Data Fig. 3).

To understand if structural transformations of Hopf solitons can also occur in other material systems, we performed stability analysis for Hopf solitons in the magnetization field $m(r)$ of chiral magnets with perpendicular BCs at confining surfaces. The micromagnetic Hamiltonian of chiral magnets resembles the Frank-Oseen elastic free energy in the isotropic elasticity limit of $K_{11} = K_{22} = K_{33} \equiv K$ in Eq. (1), with $K$ and $2\pi K/p_0$ related to the exchange and Dzyaloshinskii-Moriya interaction constants, respectively (Methods)[6,24]. We subject this chiral magnetic material to an applied magnetic field $H$ (normalized field strength $\tilde{B}$) and a uniaxial magnetocrystalline anisotropy (normalized anisotropy strength $\tilde{K}_u$), both along $\hat{z}$, which couple to $m(r)$ linearly and quadratically (Methods). We find that magnetic heliknotons can be metastable at zero $\tilde{K}_u$ and $\tilde{B}$, while finite $\tilde{K}_u$ promotes a uniform background and metastable magnetic hopfions under confinement (Fig. 2c), similar to the structural stability of Hopf solitons in LCs at $\sqrt{K_{33}/K_{22}} \approx 1$ (Fig. 2b). Through Zeeman coupling, $H$ aligns $m(r)$ linearly, and Hopf solitons can be metastable or stable for either direction of $H$ in a helical or conical background, while hopfions in a uniform background is metastable only when $H$ is in the same direction as the BC magnetization on the surfaces (along $+\hat{z}$). Outside of the stability regions of Hopf solitons, magnetic torons formed with $m(r)$ resembling the $n(r)$ of LC torons. As with previously studied magnetic Hopf solitons, the streamlines of skyrmion number density $\Omega$ (or emergent magnetic field) derived from equilibrated Hopf solitons form Hopf fibration, regardless of the embedding background or confinement conditions[17,20] (Extended Data Fig. 4).

Inspired by the numerically revealed pathway for inter-transforming LC heliknoton and hopfion, we performed simulations and experiments in a confined chiral LC to demonstrate this. Simulation results show that under confinement with perpendicular BCs for $\sqrt{K_{33}/K_{22}} \lesssim 1.1$, the helical state in $n(r)$ can transition smoothly, through conical state, to the uniform state by increasing $\tilde{E}$ (Fig. 3a). Concomitantly, a heliknoton transforms smoothly into a hopfion (Fig. 3b-d, Extended Data Fig. 5). We experimentally generated a heliknoton with laser tweezers in a confined ($d = 3p_0$) chiral LC mixture with reduced $\sqrt{K_{33}/K_{22}}$ (Fig 3e, Extended Data Fig. 6), which remained stable at no $\tilde{E}$. Upon increasing the voltage applied across the LC cell, the heliknoton-to-hopfion transformation was revealed by bright field and polarizing optical microscopies (POMs) (Fig. 3e). Notably, though hopfions and heliknotons have been known to share the same $\boldsymbol{n}(\boldsymbol{r})$ topology[9], this is the first time the two are shown as geometric embodiments of the same field topology through directly imaging their inter-transformations. This process can be paralleled with transformations between geometric shapes of objects that preserve topological invariants, like the transformation between a coffee mug and doughnut that retains the genus-one surface topology. Remarkably, this transformation is reversible when the voltage is switched off and a heliknoton can undergo a full cycle of transformation to a hopfion and back to heliknoton (Fig. 3e, Extended Data Fig. 7, Supplementary videos 2-4). A close inspection of the POM micrographs and simulated preimages of the Hopf soliton during transformation show that the full-cycle transformation process of $\boldsymbol{n}(\boldsymbol{r})$ is nonreciprocal (Fig. 3e-g).

The elapsed time needed to transform from one geometric embodiment (e.g., hopfion or heliknoton) to the other is also asymmetric, with the transformation from a hopfion to a heliknoton roughly three times longer than in the opposite direction (Fig. 3e). The full-cycle transformation is accompanied by a displacement of ~1.2 μm along the long axis of the heliknoton, or ~0.52 of the helical pitch $p_0$ of the chiral LC. These inter-transformation and displacement are reproduced quantitatively in simulations, showing that the field topology – evident by the linking number of preimages – is preserved throughout the transformation (Fig. 3f,g, Supplementary Video 5). Additionally, we identify Hopf solitons embedded in a conical background at an intermediate $\tilde{E}$ with an intermediate state of the transformation (Extended Data Fig. 5h), showing how the inter-transformation between hopfion and heliknoton progressed through an intermediate stage of a Hopf soliton in the conical background.

The heliknoton-hopfion inter-transformation and the ensuing spatial displacement repeat with periodic switching of the applied voltage on and off, leading to an activated propelling motion (Fig. 4a,b, Supplementary Video 6), similar to the "squirming" motion of 2D skyrmions[12]. This translational motion is enabled by the nonreciprocal evolution of $n(r)$ during the inter-transformations between the various geometric embodiments of the Hopf soliton with the voltage modulation within each modulation period. Experimental and computer-simulated preimage visualizations and POM micrographs, which exhibit a good agreement (Fig. 3e-g), vividly reveal these origins of the nonreciprocal structural evolution and soliton motion. Interestingly, such motion of a Hopf soliton deviates from a linear trajectory. Within each transformation cycle, the Hopf soliton undergoes a net displacement along the direction of heliknoton's long-axis orientation, while the long-axis orientation of the heliknoton with respect to $x$-axis ($\phi$) slowly changes with time (Fig. 4a-c, Extended Data Fig. 8a-d). Moreover, we found heliknotons adopt various orientations even at static equilibrium with no applied voltage and when the substrates are rubbed to achieve a uniform in-plane orientation in the background helical state at the sample mid-plane (Fig. 4d). Since the orientation of the soliton correlates with its $z$ position along $\chi_0$ in a helical background, as previously revealed by direct 3D imaging[9], this suggests that, beyond 2D lateral dynamics, our Hopf solitons also undergo orientation-correlated displacements in the third dimension along $\chi_0$. To further understand the propensity of Hopf solitons to adopt certain orientations and $z$ positions, we numerically investigated the initial ($z_i$) and final, energy-minimizing $z$ positions ($z_f$) with respect to the sample midplane before and after equilibration, as well as the final free energy (Fig. 4e-g). This was done by numerically displacing a hopfion or heliknoton from the sample midplane in their respective backgrounds before relaxing them towards equilibrium (Heliknotons so displaced were also rotated in a way consistent with the helical background; see Methods). Our results revealed that hopfions quickly relaxed and symmetrically filled up the entire vertical space between the confining substrates ($z_f = 0$), regardless of its initial position $z_i$; this yields a linear relation between $z_f - z_i$ and $z_i$ shown by the red line in Fig. 4e. On the other hand, we found the energy landscape of a heliknoton in a helical background with confinement and perpendicular BC contains multiple energy minima for spatial locations along $z$. The lowest energy minima were at $\sim 0.5 p_0$ away from the confining substrates, whereas local energy minima were distributed symmetrically around the cell midplane, leading to a nontrivial relation between $z_i$ and $z_f$ of the soliton (Fig. 4e,f). The midplane is an unstable position and corresponds to a local energy maximum. As a result, the dependence of heliknoton orientation $\phi$ on $z$ is slightly perturbed from the anticipated linear relation, though it is still monotonic and allows for the $z$ position of a heliknoton to be inferred by its orientation in POM micrographs within the same $2\pi$ rotation period (Fig. 4g). The displacement in $z$ and energy landscape are qualitatively similar for $d = 3p_0$ and $d = 4p_0$, showing these features are general for heliknotons in a helical background with confinement.

    Our analysis suggests the following scenario of periodically repeated geometric inter-transformation and activated motion. In each geometrically distinct state as a hopfion, the soliton equilibrates as it moves towards the $z = 0$ midplane in a uniform background when the voltage is on. In the subsequent heliknoton state with the voltage turned off and the background being helical, thermal fluctuations and/or sample imperfections tip it away from the then unstable $z = 0$ position towards the nearest energy minimum accompanied by a rotation. The soliton again equilibrates

towards $z = 0$ in the following hopfion state, and so on and so forth. In this process, the nonreciprocal director evolution causes 3D squirming and the Hopf soliton propels along the lateral directions while, at the same time, moves vertically between energy minima along $z$ – an effective "hopping" dynamics in the 3D space.

The squirming motion of Hopf solitons does not require complete inter-transformation between hopfion and heliknoton and their respective uniform and helical backgrounds. In thinner confined chiral LC slabs ($d = 1.7p_0$) with perpendicular BCs, Hopf solitons also propel in response to low-magnitude voltage modulation with extremes corresponding to a hopfion in a uniform background and a perturbed hopfion in a helical background (Fig. 4h-l). We found that such activated motion is characteristic for hopfions with both +1 and -1 Hopf indices (Fig. 4m,n, Extended Data Fig. 8e, Supplementary Videos 7 and 8). Numerical modeling shows the details of nonreciprocity in $\boldsymbol{n}(\boldsymbol{r})$ of temporal evolution of the Hopf soliton during a single period of voltage modulation that drives the activated motion (Fig. 4o-r).

To conclude, we have shown, for the first time, that Hopf solitons can exist as spatially localized structures without external fields or confinement in the helical background of chiral LCs within the proper range of elastic anisotropy parameters, as well as can be stable or metastable in the conical backgrounds when an external electric field is applied. The composition-based engineering of elastic constant anisotropy – energetic costs between different components of the gradient of field – provides a novel route for enhancing soliton stability, beyond the known methods of overcoming constrains of the Derrick-Hobart theorem[3,15,28]. Materials such as novel LCs with exceptionally large elastic constant anisotropy and their mixtures[30,34,35], as well as magnetic LC colloidal materials based on them[6], can be used as testbeds for stability of 3D topological solitons, expanding the selection of material systems hosting them. Moreover, we show that, by reducing the bend elastic constant, soliton stability in LCs resembles that in elastically isotropic magnets. This demonstrates elastic anisotropy engineering can further enable LCs as a testbed for general magnetic structures at a quantitative level. We have also unambiguously demonstrated that hopfions and heliknotons are geometric embodiments of the same underlying field topology and can be transformed reversibly between one another by an electric field, in a way resembling inter-transformations between geometrically different but topologically identical surfaces where the topological invariant, genus, is preserved. Moreover, in response to a modulated electric field, Hopf solitons exhibit the 3D hopping-like dynamics as a result of combined nonreciprocity in the transformation of field configurations and multi-minima energetic landscape in a confined chiral LC. The newly discovered stability of Hopf solitons and their 3D hopping dynamics is of great interest for technological applications and adds to the diversity of spatio-temporal manipulation methods of topological quasi-particles.

**Figures**

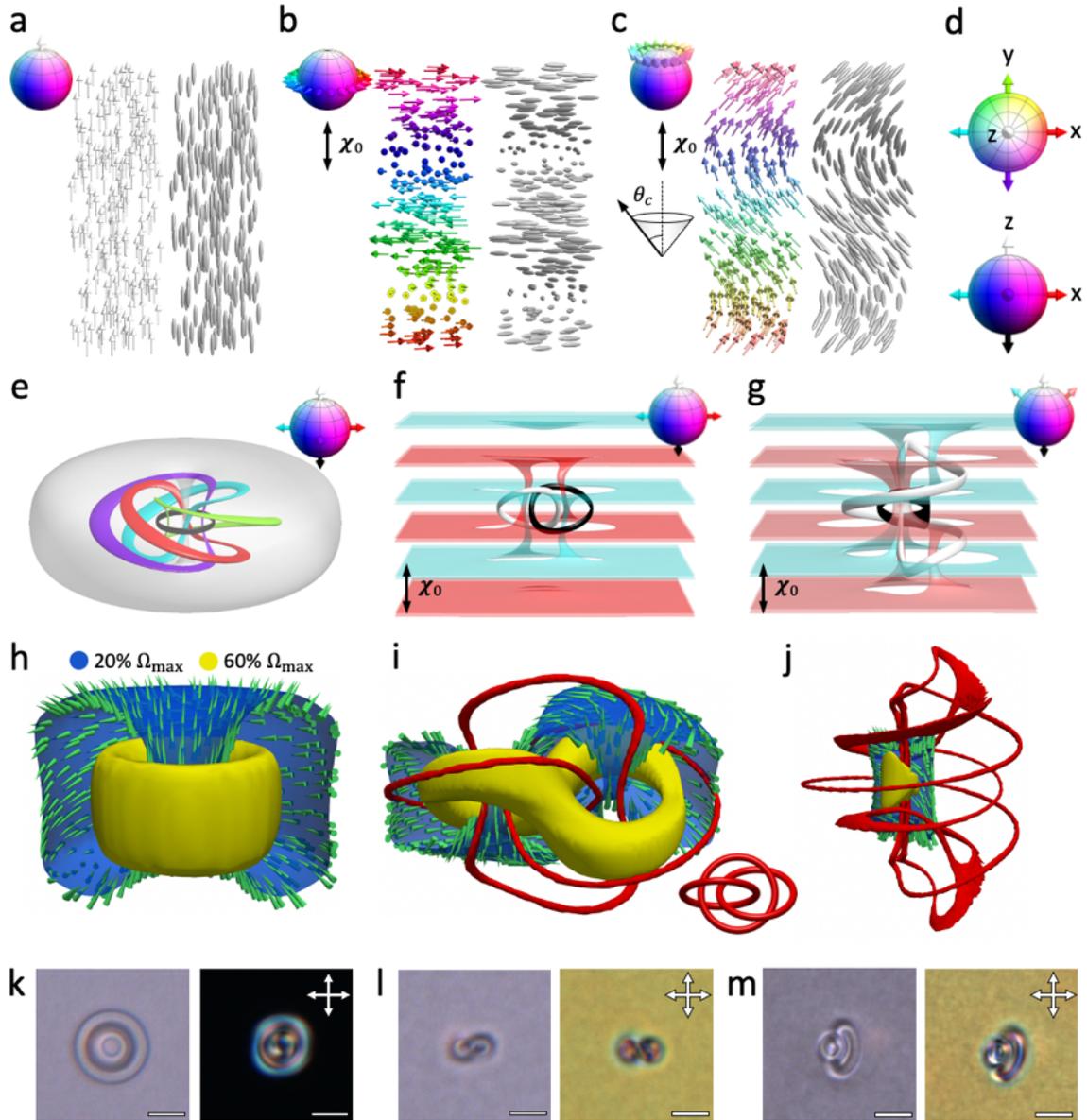

**Fig. 1 | 3D Hopf solitons in topologically trivial backgrounds**. **a-d**, Schematics of $n(r)$ backgrounds in uniform (a), helical (b), and conical (c) states represented by arrows for vector fields and ellipsoids for nonpolar director fields, respectively. The conical state is at a cone angle $\theta_c$ with respect to the helical axis $\chi_0$. Vectors are colored based on their orientations as shown in the order parameter sphere in insets and in (d). **e-g**, Preimages of vector orientations (shown in insets as arrows in the order parameter sphere) of Hopf solitons stabilized in a uniform (e), helical (f), and conical (g) background. **h-j**, Visualizations of the skyrmion number density $\Omega$ and vortex lines of stabilized Hopf solitons in different backgrounds corresponding to (e-g), respectively. The inset in (i) shows vortex line in the heliknoton can form mutually linked rings[20]. **k-m**, Hopf solitons observed in chiral LCs using bright field (left panels) and POMs (right panels) in a uniform (k), helical (l), and conical (m) background, respectively. The directions of cross polarizers are shown in the insets and scale bars are 5 μm.

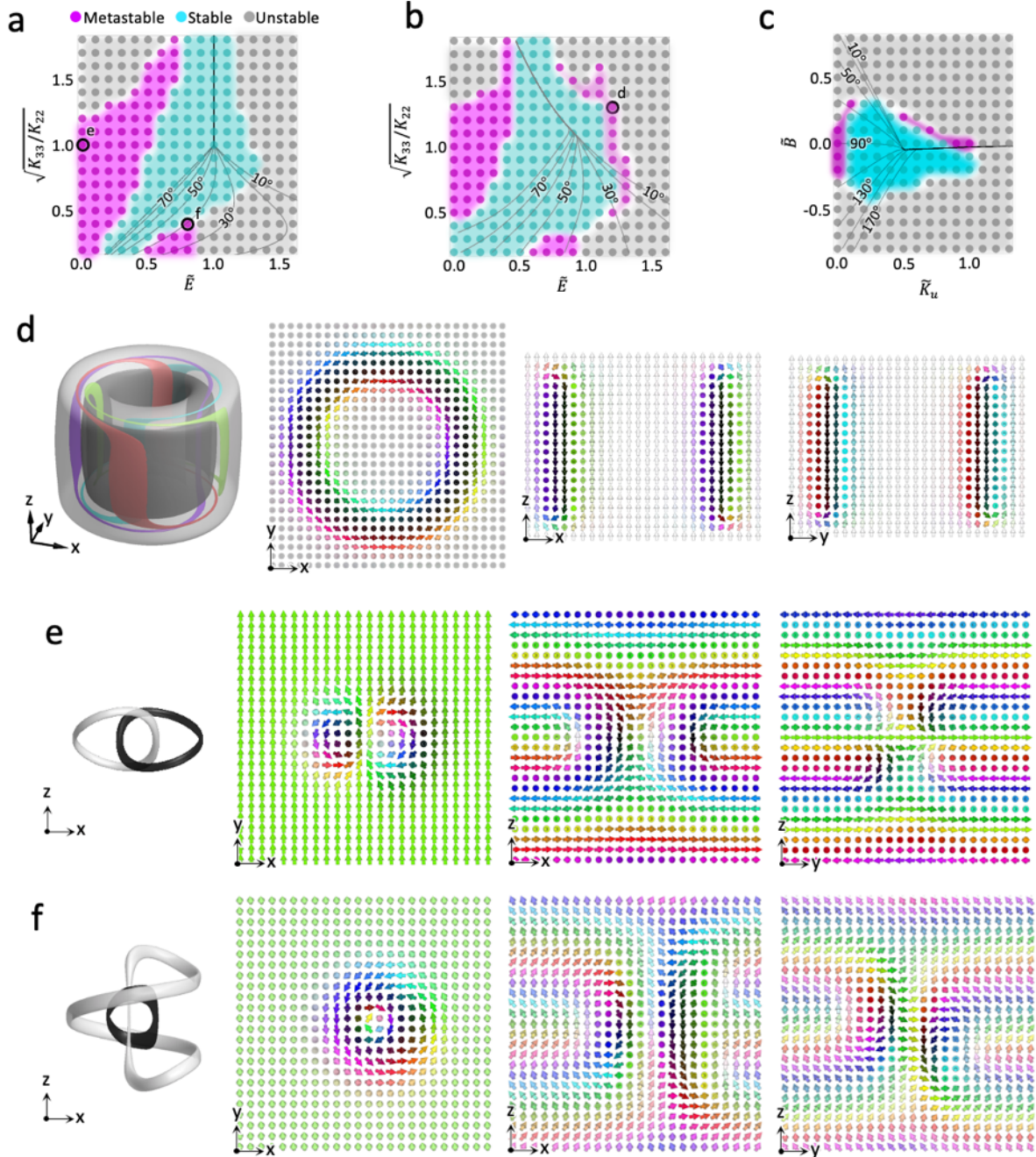

**Fig. 2 | Structural stability of Hopf solitons. a-b**, Structural stability diagrams of LC Hopf solitons in the bulk (a) and within a confined volume with perpendicular BC and $d = 3p_0$ (b). The data points are colored based on the stability of Hopf solitons, and the contour lines of background mid-plane cone angle $\theta_c$ are shown on each diagram. **c**, Structural stability diagram of magnetic Hopf solitons in a confined volume with perpendicular BC and $d = 3\lambda$. **d-f**, Preimages and director fields of representative Hopf solitons in a uniform (d), helical (e), and conical background (f), respectively. Preimages and vectors are colored based on the color scheme shown in Fig. 1d.

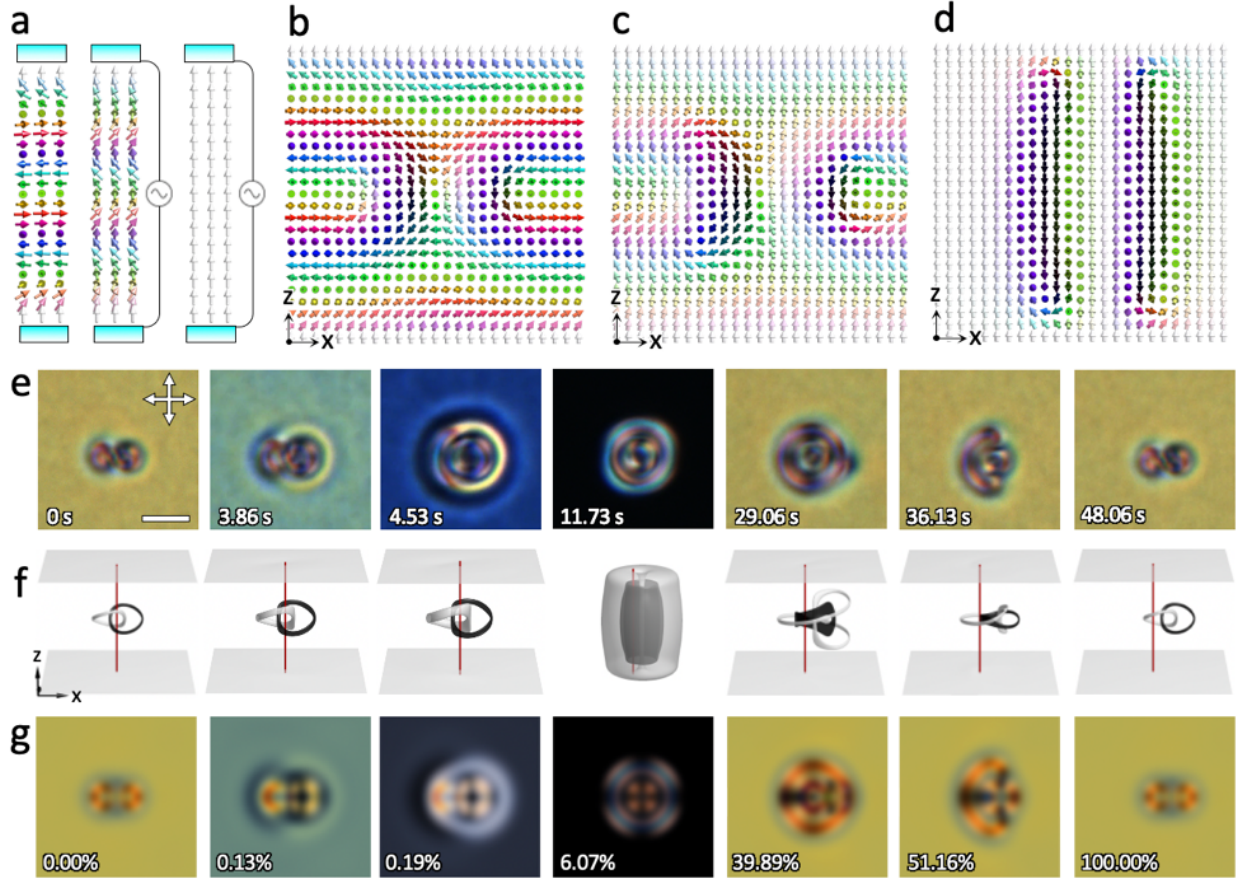

**Fig. 3 | Geometric transformation of Hopf solitons driven by electric field. a**, Schematic illustration of the experimental setup and the switching of topologically trivial backgrounds; $\theta_c$ at midplane increases from 0° to 90° as the applied voltage increases. **b-d**, Vertical midplane cross-sections through the Hopf soliton going through inter-transformation from a heliknoton to a hopfion. (b-d) correspond to simulated $\boldsymbol{n}(\boldsymbol{r})$ at 0.00%, 0.19%, 6.07% of the total simulation time in (f-g). **e**, Experimental POM snapshots of Hopf soliton switching from a heliknoton (0 s) at $U$ = 0 V to a hopfion (11.73 s) at $U$ = 3.85 V, and back to a heliknoton at $U$ = 0 V (48.06 s). $d/p_0 = 3$, $d = 7$ μm and scale bars are 5 μm. **f-g**, Simulated transformation shown by preimages of two antiparallel orientations (f; white: $+\hat{z}$, black: $-\hat{z}$) and simulated POM micrographs (g). The vertical red line going through the center of the volume in (f) is a guide to the eye. The progress of switching in numerical simulations shown as percentage of the total simulation time is labeled in each panel. In simulations, $K_{33}/K_{22} = 1$ and $d/p_0 = 3$.

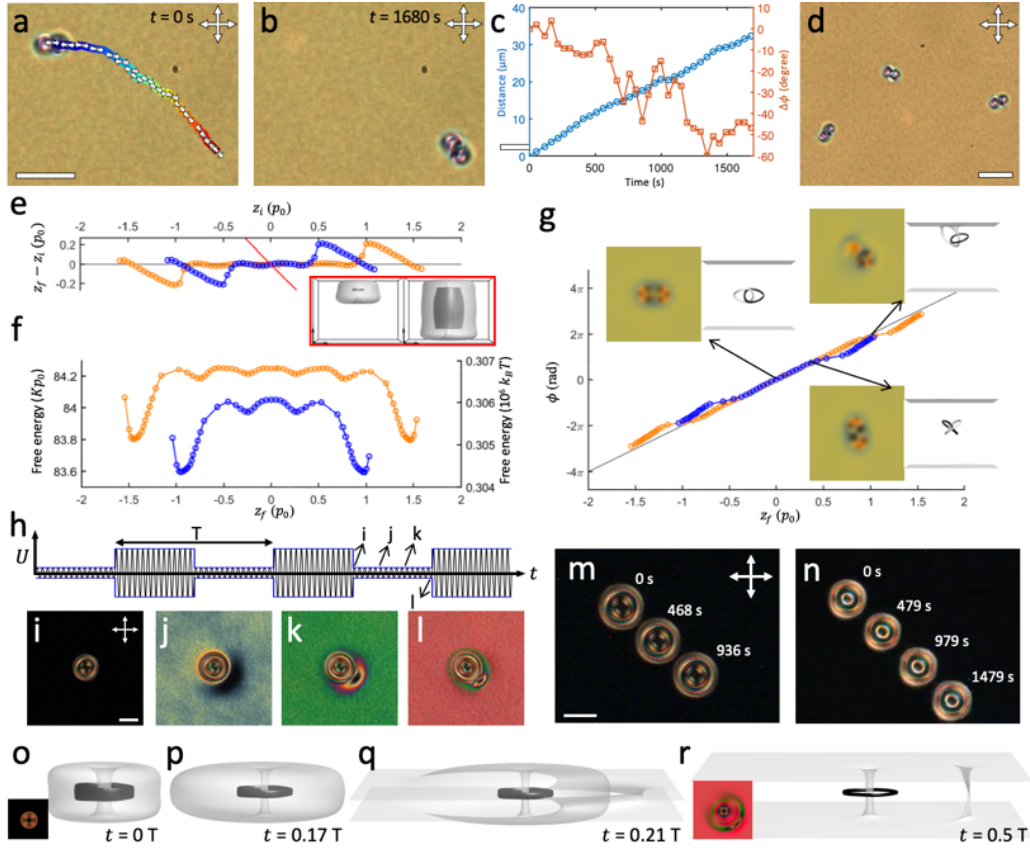

**Fig. 4 | 3D Hopping and squirming of Hopfions. a-b**, Translational and orientational displacement of a Hopfion by repeated voltage switching shown at its initial (a) and final (b) position. The 2D trajectory is color-coded by time and the long-axis orientations of the Hopf soliton in helical background at intermediate positions are shown by double arrows (plotted for every two switching cycles). **c**, Distance and change in orientation in each transformation cycle ($\Delta\phi$) of a hopping Hopf soliton shown in (b-c). **d**, Hopf solitons in different long-axis orientations in a helical background with perpendicular confinement. **e**, Hopf soliton displacement in $z$ as a function of its initial position $z_i$. **f-g**, heliknoton's free energy (f) and orientation (g) dependence on equilibrium heliknoton position $z_f$. In (e-g), heliknotons with $d = 3p_0$ and $d = 4p_0$ are shown in blue and orange. The inset and the red line in (e) correspond to hopfions. Insets in (g) show the simulated POM images (viewed along $\hat{z}$) and the preimages (viewed along $\hat{y}$) of the solitons with $z$ positions 0, $0.3p_0$, and $1p_0$ relative to the midplane (left to right) for $d = 3p_0$. The line in (g) shows $\phi = 2\pi(z_f/p_0)$ for bulk heliknotons is a guide to the eye. Free energy is in units of $Kp_0$ and $k_BT$, where $K = 6.47$pN is the average elastic constant of 5CB and $p_0 = 2.33$ μm, $k_B$ is the Boltzmann constant, and temperature $T = 300$ K. **h-l**, Snapshots of POMs of a $Q = +1$ Hopf soliton subject to background modulation (i-l) by a modulating voltage shown in (h). The modulation period T = 2 s. **m-n**, Squirming of Hopf solitons of Hopf indices $Q = +1$ (m) and $Q = −1$ (n) shown by superimposed POMs of Hopf solitons at different times. **o-r**, Field configurations of a $Q = +1$ Hopf soliton under voltage modulation in numerical simulations visualized by $\pm\hat{z}$ preimages (in white and black). Simulated POM micrographs are shown in the insets of (o) and (r). $d = 3p_0$ in experiments in (a-d) and $d = 1.7p_0$ in experiments and simulations in (i-r). Scale bars are 10 μm in (a-c) and 20 μm in (i-n).

**Methods**

**Preparation of samples**

Homogeneous mixtures of 4-Cyano-4'-pentylbiphenyl (5CB, from EM Chemicals) and 4',4''-(heptane-1,7-diyl)-dibiphenyl-4-carbonitrile (CB7CB; from SYNTHON Chemicals, Germany) were obtained by mixing the two compounds at 125°C in the isotropic phase with active stirring. The resulting mixtures at 60 or 70 wt% of 5CB were in nematic phase at room temperature[31]. The nematic mixtures were then added with a small amount of right-handed chiral additive CB-15 (EM Chemicals) to achieve right-handed chiral LC mixtures with helical pitch $p_0$ ranging from 2.33 to 10 μm as measured in a Grandjean-Cano wedge cell[36].

The LC cells were made with Indium Tin Oxide (ITO) coated glass slides or coverslips. The ITO glasses were treated with polyimide SE5661 (Nissan Chemicals) by spin-coating at 2,700 rpm for 30s and then baking at 90°C, followed by 1h at 180°C to set strong perpendicular boundary conditions for $\boldsymbol{n}(\boldsymbol{r})$ at the LC-glass interface. The polyimide-treated ITO glasses were then rubbed mildly and assembled in antiparallel rubbing directions to ensure the background $\boldsymbol{n}(\boldsymbol{r})$ is a uniform helical state with constant in-plane orientation at sample midplane at no electric field (Fig. 3a). To assemble into a cell, silica micro-cylinders with diameters 7 μm to 20 μm were used as spacers and were sandwiched between ITO glasses and fixed by UV-curable glue. Metal wires were additionally soldered to ITO glasses as electrodes for electric control. Chiral LC mixtures were then introduced into the cell by capillary forces.

To achieve electric control of LC background fields and solitons, the electrodes of the LC cells were connected to a function generator (DS345; Stanford Research Systems) operating at 1 kHz carrier frequency with sinusoidal output to preclude complex hydrodynamic effects[13].

Additionally, we used an in-house MATLAB code controlling a data acquisition board (NIDAQ-6363, National Instruments) connected to a computer for fast modulation of voltage output.

**Laser generation and imaging of Hopf solitons**

Hopf solitons were generated by holographic laser traps capable of producing predesigned patterns of laser intensity within the LC sample.[9] The tweezers setup is based on an ytterbium-doped fiber laser (YLR-10-1064, IPG Photonics, operating at 1,064 nm) and a phase-only spatial light modulator (P512-1064, Boulder Nonlinear Systems) integrated with an inverted optical microscope (IX81, Olympus)[33]. Upon focusing the 1,064 nm laser into the LC sample, local heating and optical realignment created initial $\boldsymbol{n}(\boldsymbol{r})$ that eventually relaxed into Hopf solitons under suitable energetic conditions.[9]

Bright field microscopy, polarizing optical microscopy and videomicroscopy were performed using the same IX-81 Olympus inverted microscope and a charge-coupled device camera (Flea-COL, from PointGrey Research)[33]. Phase contrast microscopy was performed using a condenser annulus and a 60X oil-immersion phase contrast objective. Differential interference contrast microscopy was performed by introducing Nomarski prisms into the light path between crossed polarizers.

Three-photon excitation fluorescence polarizing microscopy (3PEF-PM) imaging of $\boldsymbol{n}(\boldsymbol{r})$ in Hopf solitons was performed by a setup built around the same IX-81 microscope integrated with bright field microscopy and laser tweezers[9,33]. 5CB molecules in the LC mixture were excited via three-photon absorption by using a Ti-Sapphire oscillator (Chameleon Ultra II; Coherent) operating at 900 nm with 140-fs pulses at a repetition rate of 80 MHz[37]. The fluorescence signal was epi-collected by an 100X oil-immersion objective with NA = 1.44 and detected through a 417/60-nm bandpass filter by a photomultiplier tube (H5784-20, Hamamatsu). The polarization state of the excitation beam was controlled by using a polarizer and a rotatable half-wave retardation plate or a quarter-wave retardation plate. In 3PEF-PM imaging experiments, a third-order nonlinear optical process was involved, and the image intensity scales as $\cos^6\beta$, where $\beta$ is the angle between the dipole moment of the LC molecule, orientating along $\boldsymbol{n}(\boldsymbol{r})$, and the polarization of the exciting light. 3PEF-PM reveals the 3D $\boldsymbol{n}(\boldsymbol{r})$ field in LCs and were used to unambiguously confirm the geometry and topology of the observed solitonic field configurations[6,7,9,16].

Computer-simulated polarizing optical micrographs were based on the Jones matrix approach[16,33] for energy-minimizing $\boldsymbol{n}(\boldsymbol{r})$ configurations and by using the optical birefringence value of 5CB (Δn = 0.18). Computer simulations of the 3PEF-PM images were based on the $\cos^6\beta$ dependence in the image intensity, using the $\boldsymbol{n}(\boldsymbol{r})$ of Hopf solitons from energy minimizations.

**Numerical modelling**

We used numerical modelling based on energy minimization to explore the stability of Hopf solitons in chiral LCs and chiral magnets. For a chiral LC subjected to an external electric field, the total free energy density consists of Frank-Oseen elastic terms and the dielectric coupling term as shown in Eq. (1). Surface anchoring (anisotropic surface energy term accounting for preferred $\boldsymbol{n}(\boldsymbol{r})$ orientation at the surface) and saddle-splay deformation terms were not included in our modeling due to strong anchoring strength at the boundaries achieved in experiments. The material parameters were chosen to match those of 5CB except for $K_{33}$, which was a variable parameter to account for change in bend elastic constant enabled by varying the composition of LC mixture;

namely, $K_{11} = 6.4$ pN, $K_{22} = 3$ pN, and $\varepsilon_a = 13.8$ [9]. We used a normalized electric field strength $\tilde{E} = \sqrt{\frac{\varepsilon_0 \varepsilon_a}{K_{22}} \left(\frac{p_0}{2\pi}\right)^2} E$ in our modeling and structural stability diagrams such that the first-order transition between the helical and the uniform state happens at $\tilde{E} = 1$ for bulk LCs when $\sqrt{K_{33}/K_{22}} > 1$ (see below). The energy was iteratively minimized using an energy-minimization routine with finite difference discretization in space and forward Euler method in time implemented in an in-house MATLAB code[9,32]. Briefly, $\boldsymbol{n}(\boldsymbol{r})$ was updated iteratively from an initial structure using the Euler-Lagrange equation derived from Eq. (1). Relaxation was terminated when the change in the spatial average of functional derivatives, between iterations, converged and dropped below a threshold value determined for the steady-state stopping criterion, indicating an energy minimum is attained. In all simulations, the computational volume was sampled isotropically by a cubic grid at 24 gird points per $p_0$.

In bulk LC, topologically trivial background fields at different $K_{33}/K_{22}$ and $\tilde{E}$ can be derived analytically by energy-minimizing the ansatz for a general conical state twisting around $\boldsymbol{\chi}_0$ along $\hat{z}$ with variable pitch $p$ and cone angle $\theta_c$: $\boldsymbol{n}(\boldsymbol{r}) = \cos(2\pi z/p)\sin(\theta_c)\hat{x} + \sin(2\pi z/p)\sin(\theta_c)\hat{y} + \cos(\theta_c)\hat{z}$ [38]. For $\sqrt{K_{33}/K_{22}} > 1$, no conical state is stable, and a first-order phase transition boundary exists at $\tilde{E} = 1$ between the helical and uniform states. Conical states emerge when $K_{33}/K_{22} \leq 1$ and $\tilde{E} \in [\tilde{k}, 1/\tilde{k}]$ and the corresponding equilibrium pitch and cone angle are $p = p_0 \tilde{k}/\tilde{E}$ and $\cos^2(\theta_c) = \frac{1-\tilde{k}/\tilde{E}}{1-\tilde{k}^2}$, where $\tilde{k} \equiv \sqrt{K_{33}/K_{22}}$. The structural diagram of trivial background states in bulk LC is shown in Fig. 2a by $\theta_c$ contours. For the background fields of LCs in confinement, 1D simulations along z (translationally invariant in $x$ and $y$) with Dirichlet boundary conditions were performed. $\theta_c$ was measured at the midplane of the volume to yield the background $\theta_c$ contours shown in Fig 2b,c and Extended Data Figs. 1-4.

To model solitons in bulk LCs without confinement (Fig. 2a, Extend Data Fig. 1), the size of the computational volume was $4p_0$ in $x$ and $y$ and thickness $d = 10p$ in $z$. Note that since $p$ depends on $\tilde{E}$ in conical state, $d$ has to be an integral number of the equilibrium pitch $p$ (a function of $\tilde{k}$ and $\tilde{E}$) to avoid artificial frustration caused by the finite computational volume. Analytically derived $\theta_c$ was used as the boundary condition at the top and bottom surfaces. For solitons in a confined volume, Dirichlet boundary conditions of either perpendicular (along $z$, Fig. 2b,c, Extended Data Fig. 2,4) or unidirectional parallel (along $-y$, Extended Data Fig. 3) alignment at top and bottom surfaces were implemented. Periodic boundary conditions were implemented in $x$ and $y$ directions for all simulations. Two initial conditions of $\boldsymbol{n}(\boldsymbol{r})$ were constructed by either inserting a previously relaxed hopfion[7] into a uniform background with $\boldsymbol{n}_0 \parallel \hat{z}$ or a heliknoton[9] into a helical or conical background with $\boldsymbol{\chi}_0 \parallel \hat{z}$. The topology of the steady states after energy relaxation was analyzed and a data point is marked with $Q = 1$ if a Hopf soliton was stabilized with one of the initial conditions. (Fig. 2, Extended Data Fig. 1-2). Hopf index $Q$ was determined by both the linking number of preimages and numerical integral of the relaxed $\boldsymbol{n}(\boldsymbol{r})$ [9,16]. In all cases the two methods yielded consistent results except when $\boldsymbol{n}(\boldsymbol{r})$ contained singular defects and a Hopf index cannot be properly defined. The nonpolar chirality axis field $\boldsymbol{\omega}(\boldsymbol{r})$ was derived from the as relaxed soliton structure by identifying the chirality axis at all spatial coordinates with the eigenvector of the local chirality tensor $C_{ij} = n_k \epsilon_{ljk} \partial_i n_l$ [9]. The singular vortex lines in $\boldsymbol{\omega}(\boldsymbol{r})$ were determined by finding connected spatial regions where $\boldsymbol{\omega}(\boldsymbol{r})$ is ill-defined.

The algorithm for energy minimizing solitons in chiral magnets is the same as described above for chiral LCs. The micromagnetic Hamiltonian density of a chiral magnet under an external magnetic field and uniaxial magnetocrystalline anisotropy (Fig. 2c, Extend Data Fig. 4) reads[17,20]

$$f_{\text{magnet}} = \frac{J}{2}(\nabla \cdot \boldsymbol{m})^2 + D(\boldsymbol{m} \cdot \nabla \times \boldsymbol{m}) - \mu_0 M_s \boldsymbol{m} \cdot \boldsymbol{H} - K_u (\boldsymbol{m} \cdot \boldsymbol{l}_0)^2 \qquad (2)$$

with $m(r)$ the magnetization field, $J$ the exchange constant, $D$ the Dzyaloshinskii-Moriya interaction constant, $\mu_0$ the vacuum permeability, $M_s$ the saturation magnetization, $H$ the applied magnetic field, and $K_u$, $l_0$ the strength and direction of bulk uniaxial anisotropy, respectively. $H$ and $l_0$ were both along $z$. Dipolar non-local interactions were neglected for simplicity. The computational volume is parameterized by helical wavelength $\lambda \equiv 2\pi J/D$ (equivalent to $p_0$ in LCs) and Hamiltonian by dimensionless magnetic field $\tilde{B} \equiv \mu_0 \frac{M_s J}{D^2} H$ and dimensionless anisotropy strength $\tilde{K}_u \equiv \frac{J}{D^2} K_u$. The skyrmion number density $\mathbf{\Omega}$, which is proportional to emergent magnetic field in magnetic solids, was calculated as $\Omega_i = \frac{1}{8\pi} \epsilon^{ijk} m(r) \cdot (\partial_j m(r) \times \partial_k m(r))$, where $\epsilon^{ijk}$ is the totally antisymmetric tensor. Emergent fields were calculated as $(B_{\text{em}})_i = \frac{\hbar}{2} \epsilon^{ijk} m(r) \cdot (\partial_j m(r) \times \partial_k m(r))$.

To understand the energy landscape and the stability of Hopf solitons at different $z$ positions in a confined chiral LC with perpendicular BCs summarized in Fig. 4e-g, Hopf solitons were displaced from the sample midplane to different $z$ positions before relaxing towards equilibrium. For heliknotons, the $n(r)$ of a localized heliknoton was cropped away from the bulk simulation and displaced $z_i$ along $z$-axis from the midplane while rotated $2\pi(z_i/p_0)$ in the sense of the material chirality as the initial condition and relaxed at $\tilde{E} = 0$ and $\sqrt{K_{33}/K_{22}} = 1$, the same parameters used for the initial simulation of heliknoton in the bulk. The thicknesses of the simulated cells were $d = 3p_0$ or $4p_0$. After relaxation, the average position of the geometric centers of polar preimages ($n(r) = \pm \hat{z}$) was used as the final position $z_f$ of a heliknoton, and the in-plane orientation of the vector connecting the geometric centers of the polar preimages was used as heliknoton orientation $\phi$ and agrees with the long axis of a heliknoton in POM images. The geometrical centers of each polar preimage associated with the heliknoton was determined by finding the center of the minimum bounding sphere[39] of each preimage after excluding the preimages close to the top and bottom surfaces due to boundary conditions. Hopfions stabilized in a uniform background in a confined LC (perpendicular BCs) with $d = p_0$ were placed in a $d = 3p_0$ cell at different $z_i$ positions and relaxed at $\tilde{E} = 1.1$ and $\sqrt{K_{33}/K_{22}} = 1$. Free energy of the relaxed solitons was calculated by integrating Eq. (1) and the energy of a bulk helical state at $\tilde{E} = 0$ was taken as the reference.

**Additional References**

**Data availability:** All data generated or analyzed during this study are included in the published article and its Supplementary Information and are available from the corresponding author on a reasonable request.

**Code availability:** The codes used for the numerical calculations are available upon request.

**Acknowledgements:** We acknowledge discussions and technical assistance of Andrii Repula, Mariacristina Rumi, Min Shuai, Bohdan Senyuk, and Timothy White. We are grateful to Patrick Davidson for providing CB7CB used at the initial stages of this study.

**Funding:** This research was supported by the U.S. National Science Foundation grant DMR-1810513.

**Author Contributions:** J.-S.B. T. performed experiments. J.-S.B. T. and J.-S.W. performed numerical modelling. I.I.S. directed research and provided funding. J.-S.B. T., J.-S.W. and I.I.S. wrote the manuscript.

**Competing interests:** The authors have no competing interests.

**Additional Information:** Supplementary Information is available for this paper. This file contains the Supplementary Information which includes Supplementary Videos 1-8.

**Extended Data Figures**

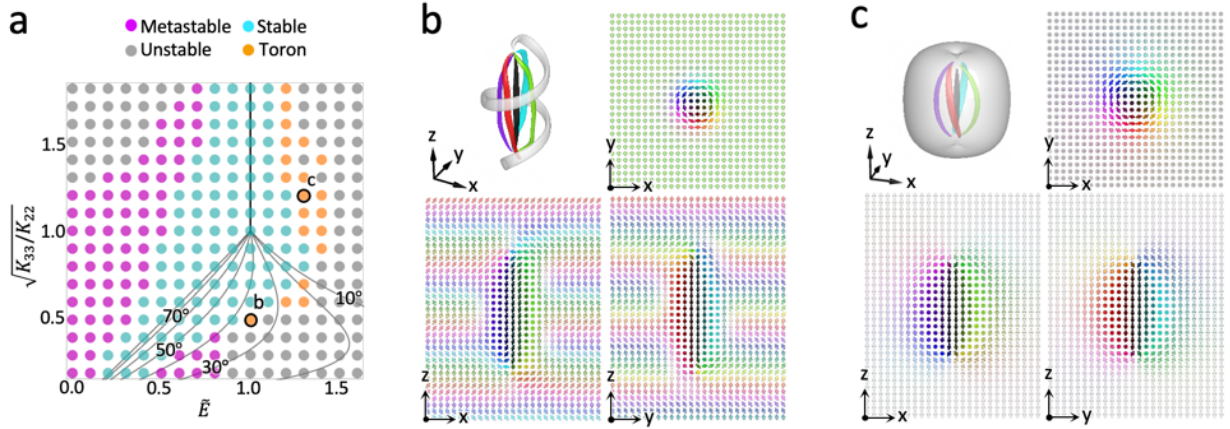

**Extended Data Fig. 1 | Numerical simulations of Hopf solitons in the chiral LC bulk. a**, Structural stability diagram of Hopf solitons in the chiral LC bulk and parameter regions showing stability of torons. The $\theta_c$ contour lines of the embedding background are shown on the diagram. **b-c**, Preimages and $\boldsymbol{n}(\boldsymbol{r})$ in different cross-sections of torons in a conical background (b) and a uniform background (c) with corresponding parameters labelled in (a).

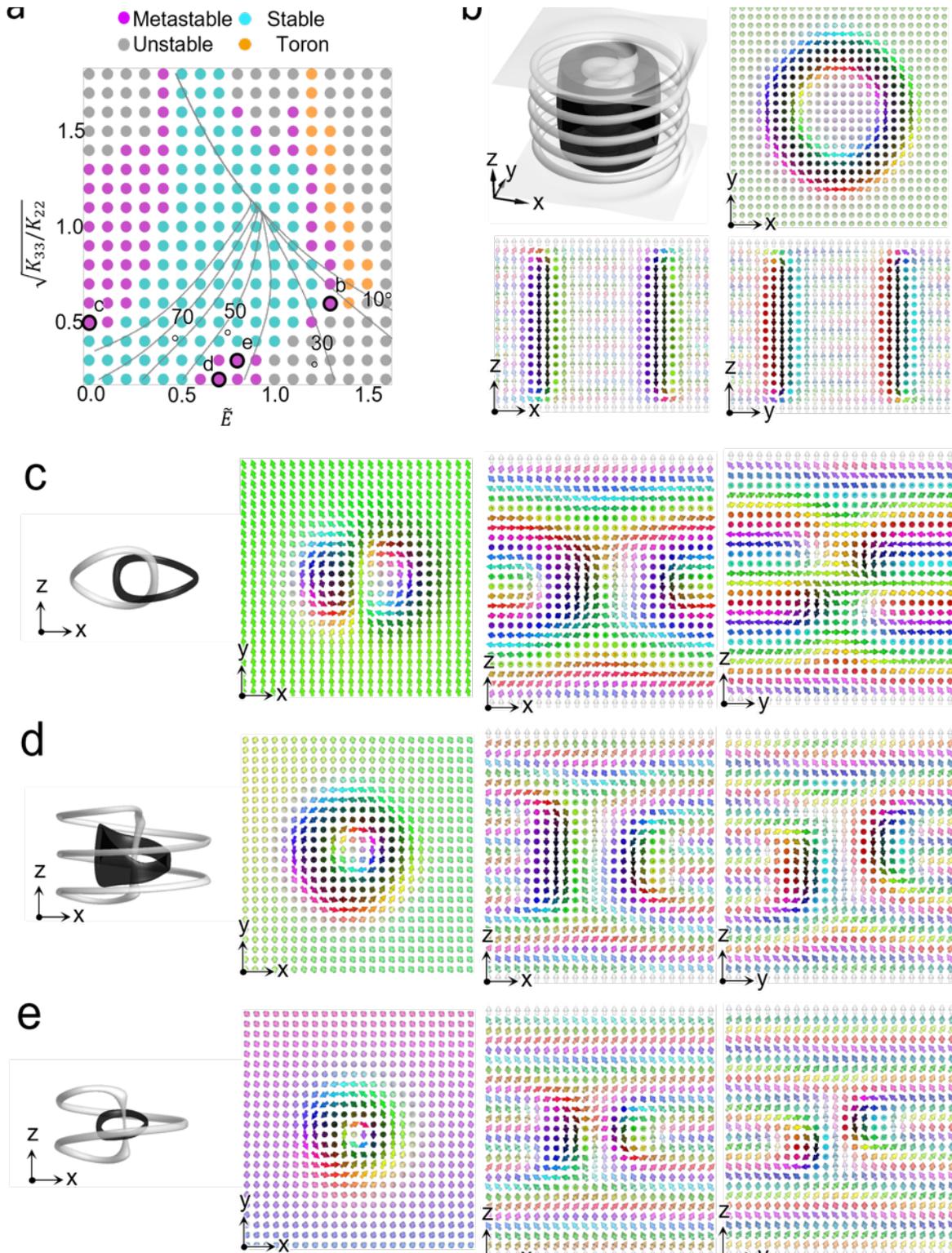

**Extended Data Fig. 2 | Numerical simulations of Hopf solitons in chiral LCs with confinement and perpendicular BC. a**, Structural stability diagrams of Hopf solitons in a confined chiral LC with perpendicular BC. Colors represent the stability of Hopf solitons, with torons distinguished from smooth topologically trivial backgrounds. **b-e**, Preimages and $n(r)$ in different cross-sections of Hopf solitons with corresponding parameters labeled in the stability diagrams in (a).

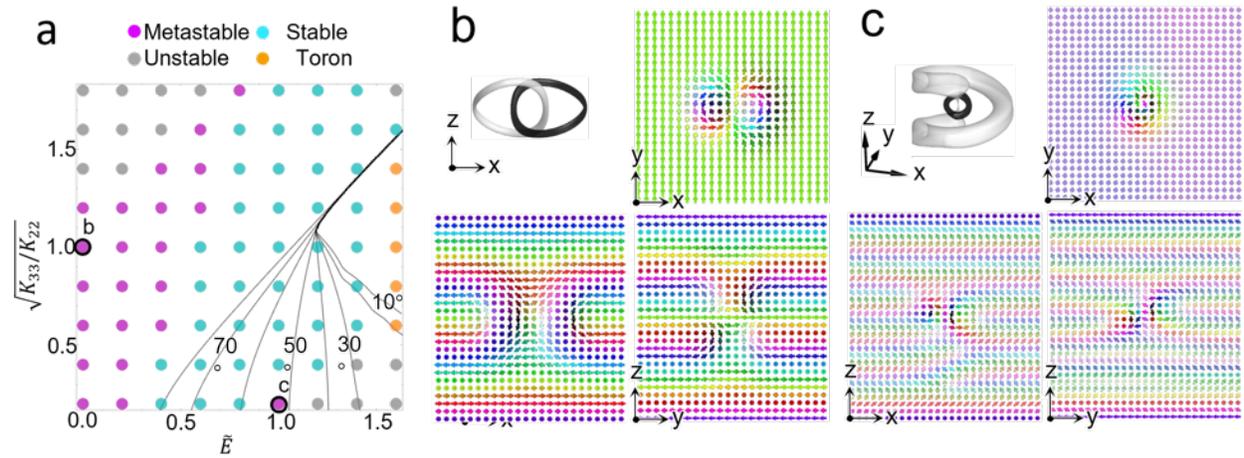

**Extended Data Fig. 3 | Numerical simulations of Hopf solitons in chiral LCs with confinement and parallel BC. a**, Structural stability diagrams of Hopf solitons in a chiral LC with confinement and parallel BC. Colors represent the stability of Hopf solitons, with torons distinguished from smooth topologically trivial backgrounds. **b-c**, Preimages and $n(r)$ in different cross-sections of Hopf solitons with corresponding parameters labelled in the stability diagrams in (a).

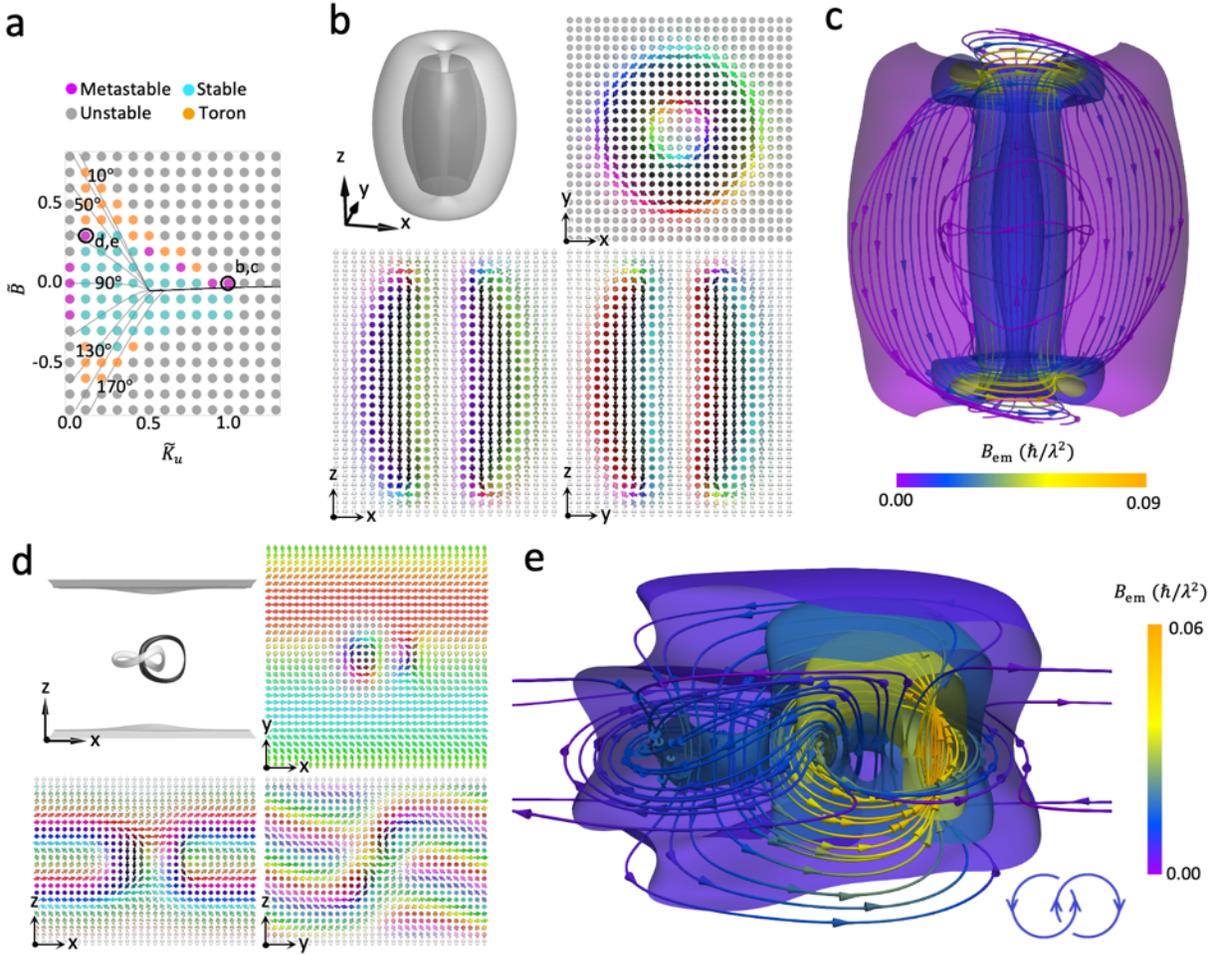

**Extended Data Fig. 4 | Numerical simulations of Hopf solitons in confined chiral magnets. a**, Structural stability diagram of Hopf solitons in a confined chiral magnet with perpendicular BCs achieved by interfacial perpendicular magnetic anisotropy[17]. Colors represent the stability of Hopf solitons, with torons distinguished from smooth topologically trivial backgrounds. **b-c**, Preimages and $m(r)$ in different cross-sections (b) and the emergent field shown by streamlines and magnitude isosurfaces (c) of a hopfion with parameters labelled in (a). Cones on the streamlines indicate the local orientation of the emergent field. **d-e**, Similar visualizations of a heliknoton with parameters labelled in (a). The inset in (e) shows each pair of closed streamline has a linking number +1.

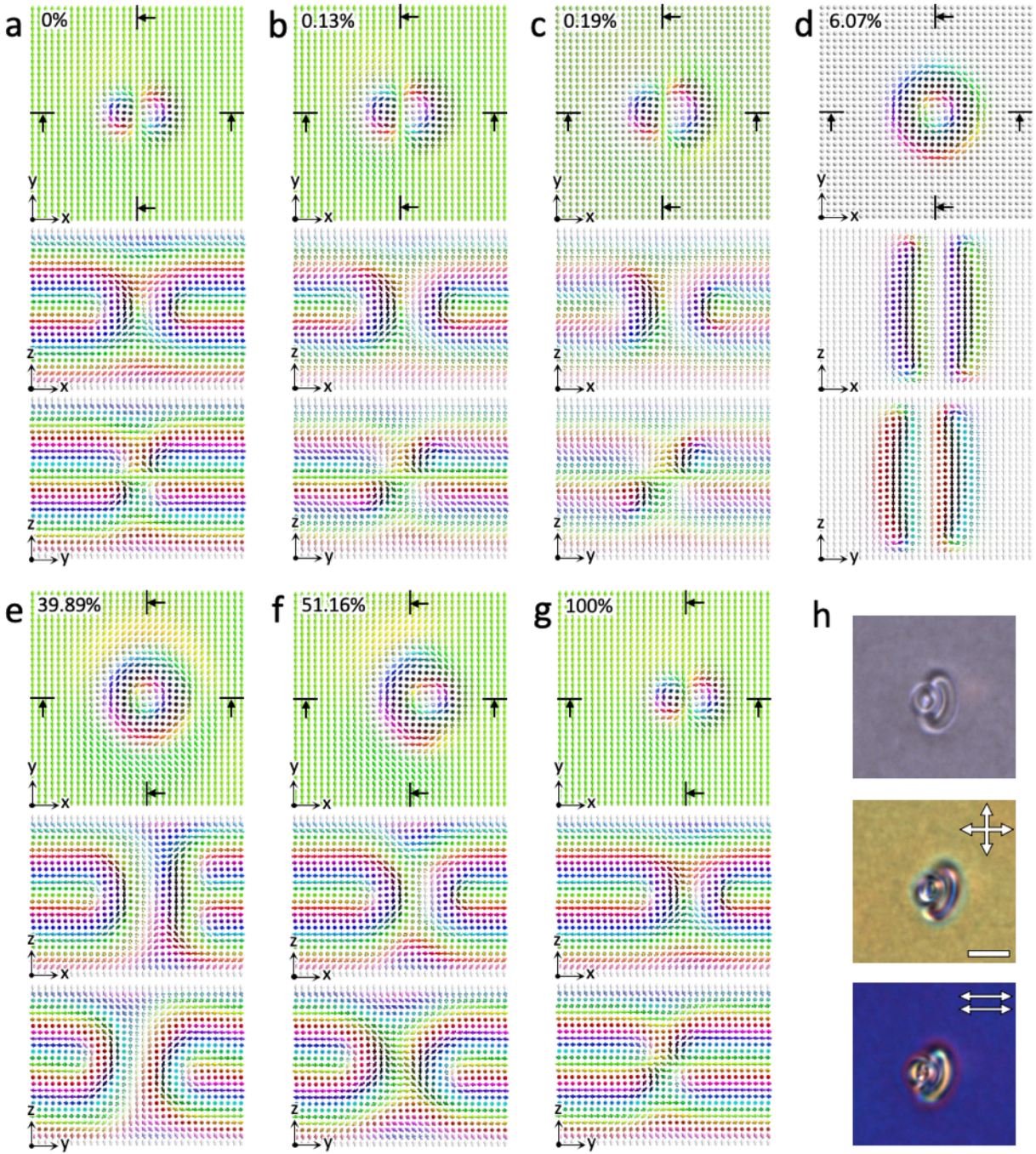

**Extended Data Fig. 5 | Simulations of the geometric inter-transformation of a Hopf soliton and the intermediate state in a conical background.** The structures correspond to those shown in Fig. 3. **a-g**, $n(r)$ of a transforming Hopf soliton corresponding to Fig. 3f,g shown in orthogonal cross-sections. The $xy$-cross-sections go through the midplane of the computational volume and the positions of vertical cross-sections ($xz$ and $yz$) are labeled. **h**, A Hopf soliton stabilized in the conical background at $U = 1.76$ V shown in brightfield and POMs (crossed and parallel polarizers; labeled top right). $d = 3p_0$, $d = 7$ μm and scale bar is 5 μm.

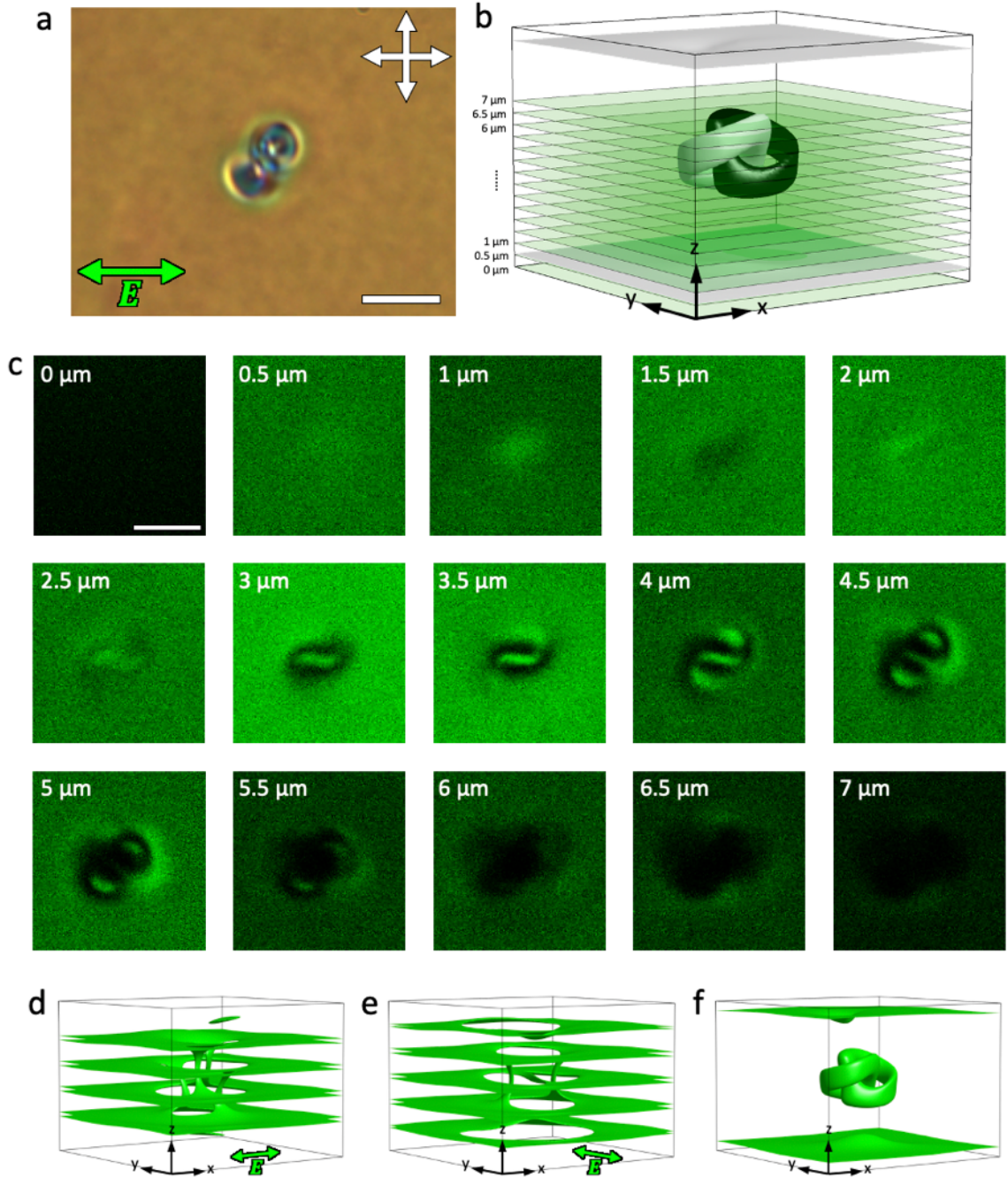

**Extended Data Fig. 6 | Three-photon excitation fluorescence polarizing microscopy (3PEF-PM) images of a Hopf soliton. a**, Polarizing micrograph of a Hopf soliton in a helical background with confinement and perpendicular BC. **b**. Schematic illustration of the vertical sectioning of 3PEF-PM images through a Hopf soliton shown by preimages of $\pm\hat{z}$ vector orientations. **c**, 3PEF-PM images of a Hopf soliton in horizontal cross-sections shown in (b) with the polarization direction of the excitation light shown in the lower-left of (a). **d-f**, Simulated 3PEF-PM images of a Hopf soliton in a helical background excited by linearly polarized (d & e, polarizations labeled) and circularly polarized (f) femtosecond laser illumination. $d = 3p_0$, $d = 7$ μm and scale bars are 5 μm.

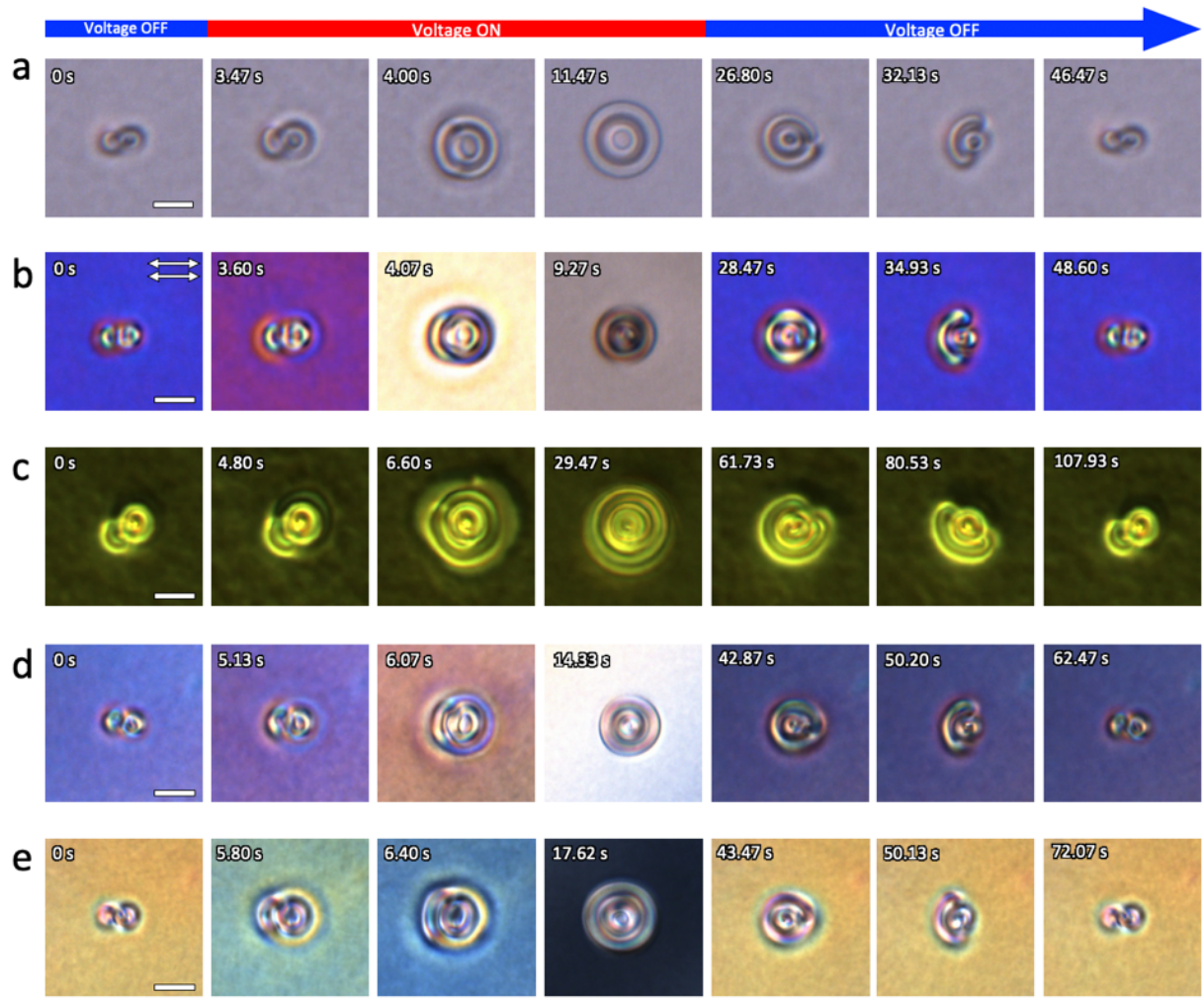

**Extended Data Fig. 7 | Geometrical inter-transformation of Hopf solitons observed by different optical imaging modalities.** The structures correspond to those shown in Fig. 3. **a**, Brightfield imaging. **b**, POM (parallel polarizers). **c**, Phase contrast microscopy. **d-e**, Differential interference contrast microscopy images obtained for two different Nomarski prism positions. Scale bars are 5 µm.

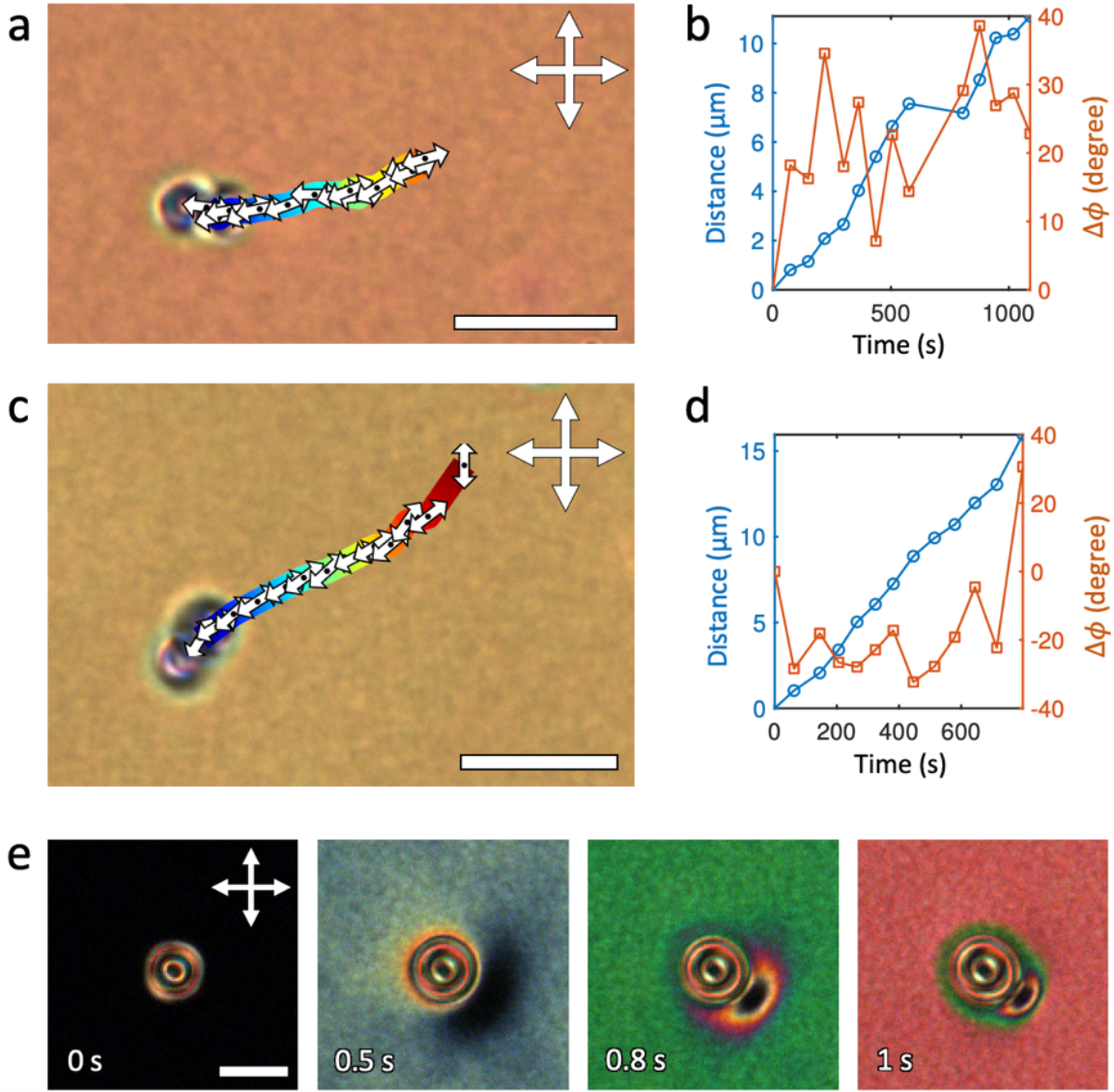

**Extended Data Fig. 8 | Additional details of hopping and propelling of Hopf solitons. a-d**, Extra data of hopping of Hopf solitons by electric switching shown by the POMs of solitons at its initial position with 2D color-coded trajectories and orientations shown by double arrows (a,c) and the corresponding distance and change in orientation in each transformation cycle ($\Delta\phi$) (b,d). **e**, Snapshots of POMs of a $Q = -1$ Hopf soliton subject to background modulation by a modulating applied voltage profile shown in Fig. 4h. Scale bar 10 μm.

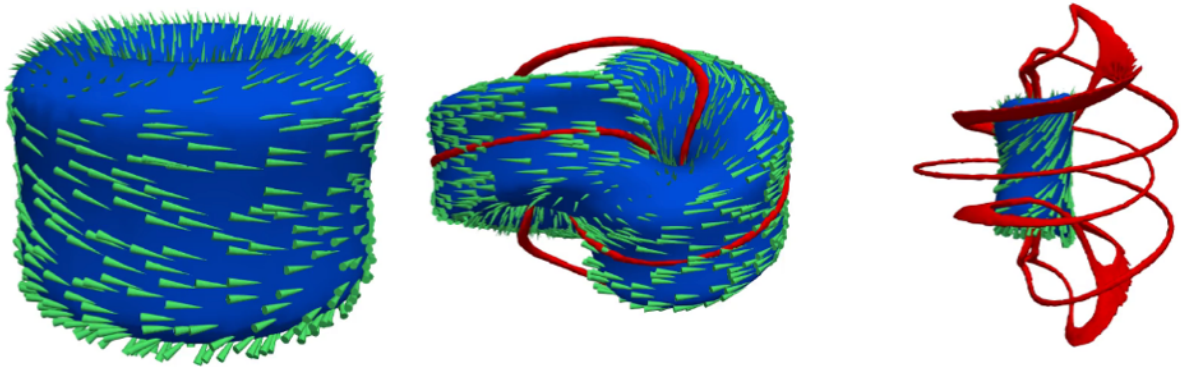

**Supplementary Video 1** | Visualizations of the skyrmion number density (blue isosurfaces and green arrows) and vortex lines in the chirality axis field (red) of numerically simulated Hopf solitons in different backgrounds, using the same structures as shown in Fig. 1h-j.

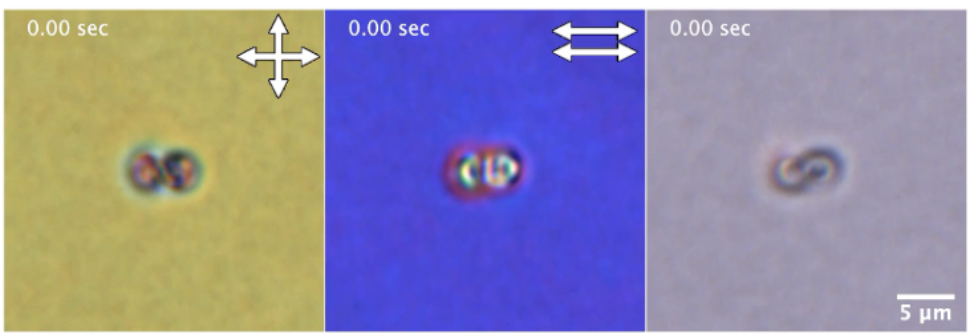

**Supplementary Video 2** | Geometric inter-transformation of a Hopf soliton observed by POM videos with crossed polarizers and parallel polarizers (left and middle) and bright-field microscopy video (right).

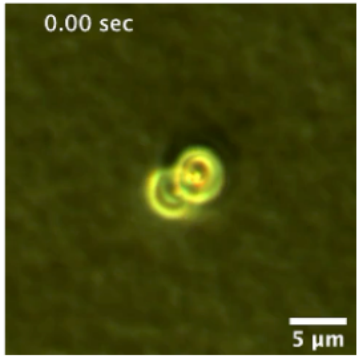

**Supplementary Video 3** | Geometric inter-transformation of a Hopf soliton observed by phase contrast microscopy.

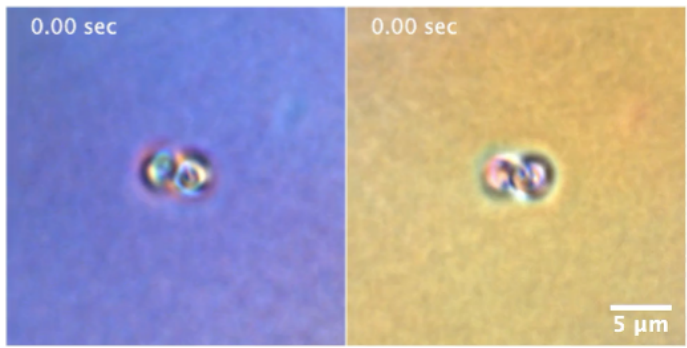

**Supplementary Video 4** | Geometric inter-transformation of a Hopf soliton observed by differential interference contrast microscopy with two different Nomarski prism positions.

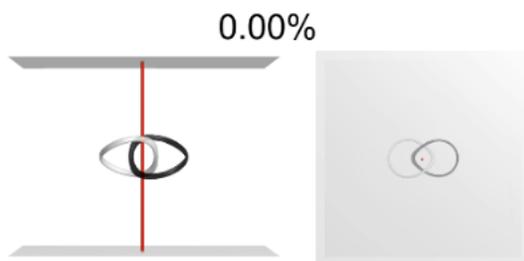

**Supplementary Video 5** | Simulated geometric inter-transformation of a Hopf soliton. The Hopf soliton is visualized by preimages of two antiparallel vector orientations ($\hat{z}$ in white and $-\hat{z}$ in black), and the vertical redline passing through the center of the initial position of the soliton serves as a guide to the eye.

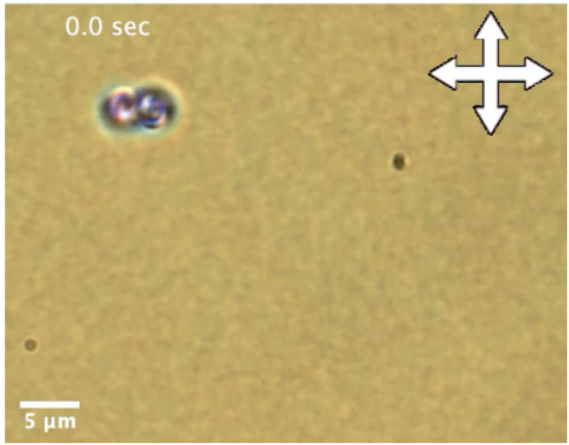

**Supplementary Video 6 |** Hopping of a Hopf soliton through repeated inter-transformation between a heliknoton and a hopfion. The voltage was switched between 0 and 3.85 V.

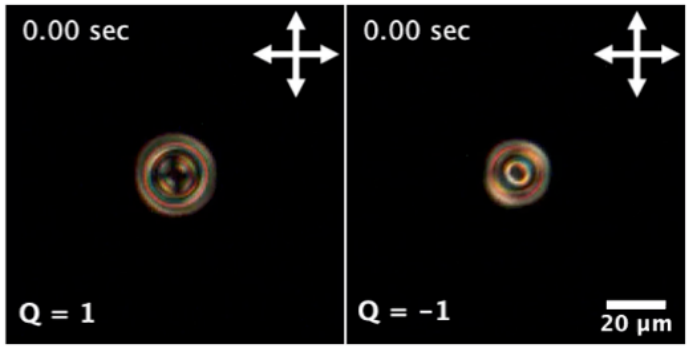

**Supplementary Video 7 |** Hopf solitons with +1 and -1 Hopf indices evolving under one full period of voltage modulation. The voltage amplitude varies between 2.27 and 0.07 V with a period of 2 s. Here $d = 1.7p_0$ and $d = 10$ μm.

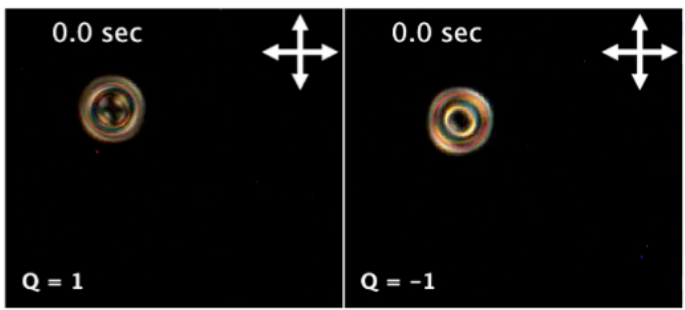

**Supplementary Video 8 |** Squirming motion of Hopf solitons with +1 and -1 Hopf indices under repeated voltage modulations. The voltage amplitude varies between 2.27 and 0.07 V with a period of 2 s. Here $d = 1.7p_0$ and $d = 10$ μm.